\documentstyle[11pt,aaspp4]{article}

\def\etal	{{et al.}}

\begin{document}
\slugcomment{Scheduled for ApJS, Jun 2000}

\title{Calibrating Array Detectors}
\author{
D.~J.~Fixsen\altaffilmark{1}$^,$\altaffilmark{3}, 
S.~H.~Moseley\altaffilmark{2}, and  
R.~G.~Arendt\altaffilmark{1}}

\altaffiltext{1}{Raytheon-ITSS Corp., Code 685, NASA/GSFC, Greenbelt, MD 20771}
\altaffiltext{2}{Code 685, Infrared Astrophysics Branch, Goddard Space Flight 
Center, Greenbelt, MD 20771}
\altaffiltext{3}{email address fixsen@stars.gsfc.nasa.gov}

\begin{abstract}
The development of sensitive large format imaging arrays for the
infrared promises to provide revolutionary capabilities for space
astronomy. For example, the Infrared Array Camera (IRAC) on SIRTF will
use four $256\times 256$ arrays to provide background limited high spatial
resolution images of the sky in the 3 to 8 $\mu$m spectral region. In order to
reach the performance limits possible with this generation of sensitive
detectors, calibration procedures must be developed so that uncertainties
in detector calibration will always be dominated by photon statistics
from the dark sky as a major system noise source. In the near infrared,
where the faint extragalactic sky is observed through the scattered and
reemitted zodiacal light from our solar system, calibration is particularly
important. Faint sources must be detected on this brighter local
foreground.

We present a procedure for calibrating imaging systems and
analyzing such data. In our approach, by proper choice of observing
strategy, information about detector parameters is encoded in the sky
measurements. Proper analysis allows us to simultaneously solve for sky
brightness and detector parameters, and provides accurate formal error
estimates.

This approach allows us to extract the calibration from the observations 
themselves; little
or no additional information is necessary to allow full interpretation of
the data. Further, this approach allows refinement and verification of
detector parameters during the mission, and thus does not depend on 
{\it a priori} knowledge of the system or ground calibration for 
interpretation of images.

\end{abstract}
\section{Introduction}

The Infrared Array Camera (IRAC) (Fazio \etal\ 1998) will employ four 
$256\times256$ 
imaging infrared arrays and the cooled telescope of the SIRTF to produce 
images of the sky which are limited by the photon statistics from the natural 
background, which, in this spectral region (8-25 $\mu$m), is dominated by 
scattered and emitted light from the zodiacal dust particles. This will be 
typical of future applications of infrared detectors in space. 
In order to produce high quality images in the
presence of this strong background, the relative response of the different
pixels in the detector array must be known to high precision. A technique
must be developed that allows the detector properties to be determined in
operation, so that the requisite stability can be experimentally verified,
and changes in response can be measured and included in the analysis of
the data. We present a technique by which the detector properties
are determined simultaneously with the estimates of sky brightness, and
formal errors developed for both instrument and sky parameters. 

We observe the same area of the sky with the detector array 
at a number of spatially offset positions. These observations are used to 
set up a system of linear equations involving both sky brightness and detector
properties. In solving this system of equations, we can deduce the sky
brightness and detector gain and offset parameters. By appropriate
choices of offset spacings and sky brightness distributions, this
technique allows us to continuously improve our knowledge of the detector
properties or detect changes. This approach embeds the
relative calibration of the detector array into the survey process; all
information required to produce an internally consistent survey
can deduced from the survey itself. Since the data on which the calibration
is based is the survey itself, it is the way to calibrate the data which is, 
in some sense, least susceptible to systematic errors. In the case that an 
{\it a priori} calibration is used, this technique offers a method to test
internal consistency.

In this paper, we describe this least squares solution for sky and
detector properties, and suggest implementations of the technique for the
IRAC instrument. We present the analysis of synthetic Wide-Field
Infrared Explorer (WIRE) data and real Hubble NICMOS data, in
which we derive the sky brightness, detector gain and detector offset. 
(We had planned a demonstration of the technique on the Wide-field
Infrared Explorer, but its unfortunate demise renders the point moot.)
The results are encouraging, and form the basis of our plans for the
analysis of the IRAC imaging data. Optimization of the observational strategy 
to produce the best encoding of the detector parameters in the survey
observations is treated in a separate paper (Arendt, Fixsen \& Moseley 2000).
This approach can offer significant insurance to the observer, in that
regardless of the availability or applicability of independent relative
calibration data for the instrument, sufficient information is present in
the observations themselves to allow the relative calibration of the
data. This provides the capability for the observer to validate the
statistical properties of the data or to calibrate it as required.

Least squares techniques are an important staple of model fitting.
In this paper, we use a least squares technique, combined
with sampling over a wide range of spatial scales, to produce an intensity
calibration for the imaging system. Investigators have long used "sky flats" 
to produce estimates of system response (e.g. Joyce, 1992). In this process, 
images are taken at a variety of positions around the object of interest.
These images are often processed using median filtering to produce estimates
of detector response. In this paper , we derive the full algorithm for
optimal use of the sampled data for intensity calibration of an imaging
detector. This algorithm then allows us to ask more sophisticated
questions important for planning observations, such as comparing the
relative goodness of different sampling procedures (Arendt et al. 1999).
This algorithm provides the optimal tool for calibrating imaging detectors;
if the algorithm does not produce reliable results, it is indicative of
incompleteness in the sampling of the sky. The algorithm provides an
optimal detector calibration based on the data provided it. If a priori
information about the detector is known, the algorithm can be adjusted to
include it.

Other algorithms have been described in the literature for
analyzing dithered image data. The drizzle method (Fruchter and Hook, 1998)
is an approach for combining undersampled dithered images to produce a
single combined image with improved resolution and signal-to-noise ratio.
However, this technique is a means of a producing final image from calibrated
data, and is not intended as a method of deriving the detector calibration.

Future observatories will generate survey data. The accuracy of analysis of
these data will depend on a
clear understanding of the statistical properties of the uncertainties in
the data, their level, and spatial and temporal correlations. We present
an approach for the analysis of such data, with specific application to the
imaging data from the SIRTF IRAC instrument.

This comprehensive least squares approach has been successfully applied to
the analysis of the data from the FIRAS instrument on COBE, in which a 
complex instrument model was required (Fixsen \etal 1994).

\section{Overview}
The following equations show the derivation of the simultaneous extraction 
of sky brightness and instrument parameters to the data. The advantages of 
this system are: 1) It uses the
same data for calibration and observation which saves separate observation
time for
calibration and uses the same time and exactly the same conditions for
calibration and observation. 2) It uses a well understood process for
calibration allowing for complete error analysis and flexible response
in the case that unexpected errors arise. 3) It {\it explicitly} includes
the uncertainties and correlations introduced in the calibration process
in the uncertainties of the resulting data. We
focus on an imaging array observing sections of the sky, but the derivation
is either directly applicable or easily generalized to other problems.

The underlying process is a simple linear fit which is easily understood,
although the matrices involved are unwieldy. The inverses
of the matrices are assumed to exist. If there are problems inverting
these matrices, it is an indication that 
information is missing in the calibration process. We do not go into
detail about the convergence or singularities of the process, but these need
to be addressed as they show key weaknesses in the
calibration process and can generally be corrected by improving the 
measurement strategy (Arendt et al. 2000).

Since the details of the calibration process leave their impact on the noise
characteristics of the final data set, the procedure for taking data must
be be carefully designed. This is not unique to this particular process for 
calibration, but this procedure
makes the costs of poor measurement strategies obvious.

\section{Derivation of the Algorithm}
We follow the Einstein summation convention
and use different indices for the different vector 
spaces. Latin indices are used for the raw data and instrument pixels while 
greek indices are used for the derived solution and the sky pixels. We use 
the same variable names for the contravariant and covariant cases even though 
the numerical values are different, because the underlying information is 
the same (see Table 1).

Consider the general solution, where we have a model of the data, 
$H^i(\theta^\mu)$, where $\theta^\mu$ is a vector of parameters which 
includes both detector and sky parameters.
First we linearize the equation, about a point $\Theta^\mu$ at or near the 
solution yielding:
\begin{equation}
H^i(\theta^\mu)\approx H^i(\Theta^\mu)+H^i_\mu \delta^\mu,
\end{equation}
where $H^i_\mu=\partial H^i/\partial\theta^\mu$. The derivatives are performed 
at $\Theta^\mu$ and $\delta^\mu$ are perturbations from $\Theta^\mu$
($\delta^\mu=\theta^\mu-\Theta^\mu$).

Given a data set $D^i$ we define $\Delta^i=D^i-H^i(\Theta^\mu)$.
With a symmetric weight matrix, $W_{ij}$, $\chi^2$ is calculated as
\begin{equation}
\chi^2 = (\Delta^i-H^i_\mu\delta^{\mu})W_{ij}(\Delta^j-H^j_\nu\delta^\nu)
\end{equation}
and its minimum is determined by
\begin{eqnarray}
\frac{\partial\chi^2}{\partial\delta^\omega} &=& -H^i_\omega W_{ij}
(\Delta^j-H^j_\nu\delta^\nu)-(\Delta^i-H^i_\mu\delta^\mu)W_{ij}H^j_\omega
\nonumber \\
&=& -2 H^i_\omega W_{ij}\Delta^j+ 2 H^i_\omega W_{ij}H^i_\nu\delta^\nu=0.
\end{eqnarray}
Thus the solution for $\delta^\mu$ can be expressed as
\begin{equation}
\delta^\mu=(H^i_\mu W_{ij} H^j_\nu)^{-1} H_\nu ^k W_{kl} \Delta^l=(H^i_\mu
H_{i\nu})^{-1}H^k_\nu\Delta_k.
\end{equation}

There are several potential pitfalls here particularly if the second derivative,
$H^i_{\mu\nu}=\partial H^i_\mu/\partial\theta^\nu$, is ill-behaved in the region of 
interest. If $H^i_{\mu\nu}\delta^\mu\delta^\nu>1$ the expansion point $\Theta^\mu$
is too far from the solution.  A new $\Theta^\mu$ closer to the solution should be 
used. If $H^i_{\mu\nu}(H^i_\mu H_{i\nu})^{-1}$ is close to 1 or larger 
a full differential geometric treatment is in order which is beyond the
scope of this paper.

The inversion of the matrix $H^i_\mu H_{i\nu}$ is the hard part of 
the problem. In what follows we show how properties of this matrix 
that frequently exist can reduce the problem to one that can be computed on a
modest computer.  The inverse of the matrix is also the covariance matrix
of the parameters including the sky parameters.

It is also interesting that:
\begin{equation}
\delta_\mu = H^i_\mu \Delta_i.
\end{equation}
This is a simple mnemonic to remember the solution. It also shows that the 
covariant form of the solution on the left is like the covariant form of the 
data on the right.  This is the weighted form of the solution needed if one 
desires to fit this solution to some higher level theory.  This can be
done even if the matrix cannot be inverted.

To develop a more tractable form of equation (4),
we separate the detector parameters from the sky parameters. 
\begin{equation}
\delta^\mu=(X^1 \dots X^{\cal P},\delta S^1 \dots \delta S^\Gamma).
\end{equation}
The parameters are not required to have the same units; the weight
matrix has all of the appropriate inverse units. 
Analogously the parameter weight matrix is separated into 3 parts,
\begin{equation}
H^i_\mu W_{ij}H^j_\nu=\left(
\begin{array}{ll}
 A & B \\
 B^T & C
\end{array}
\right).
\end{equation}

The part dealing with the instrument is $A=H^i_q W_{ij}H^j_r$. The part 
dealing with the resulting sky map is $C=H^i_\alpha W_{ij}H^j_\beta$. And 
the connections between them are $B=H^i_\alpha W_{ij}H^j_q$. The 
covariance matrix (inverse of the weight matrix) is broken into
the same sorts of parts. Often, each detector observes only one sky pixel at 
a time and the weight matrix is simple enough that the large submatrix, 
$C=H^i_\alpha W_{ij} H^j_\beta$ can be easily stored and inverted. 
Let us then consider 
\begin{equation}
(H^i_\mu W_{ij}H^j_\nu)^{-1}=(H^i_\mu~H_{i\nu})^{-1}=\left(
\begin{array}{ll}
 Q & R \\
 R^T & \Psi
\end{array}
\right).
\end{equation}

The inverse or covariance can be calculated by:
\begin{equation}
Q=(A-B C^{-1} B^T)^{-1}
\end{equation}
\begin{equation}
R = -Q B C^{-1}
\end{equation}
and
\begin{equation}
\Psi= C^{-1} +C^{-1}B^TQBC^{-1}.
\end{equation}

When the only interest is in the uncertainties in the array parameters,
(e.g. when the calibration is used for other data) only $Q$ is needed.
Similarly, if only the sky uncertainties are required, only $\Psi$ is needed.

The covariance of the derived sky, $\Psi$, is composed of two parts. The 
$C^{-1}$ is the direct propagation of the measurement errors to the sky. The
other part $C^{-1}B^TQBC^{-1}$ shows the additional uncertainty due to 
the calibration. For a well chosen set of observations this part can
approach $({\cal P}/PM)C^{-1}$, the limit set by the number statistics.

The matrix, $Q$, is much smaller than $H^i_\mu H_{i\nu}$, but still may be
inconveniently large.
Equation (4) is really a system of linear equations. By substituting 
equation (8) into equation (4) and retaining only the first ${\cal P}$ equations 
we have:
\begin{equation}
X= (Q H^i_q + R H^i_\alpha)\Delta_i = Q Y
\end{equation}
where
\begin{equation}
Y = H^i_q\Delta_i- B C^{-1} H^i_\alpha\Delta_i.
\end{equation}

The matrix $A$ relates the detector parameters to each other. With care these 
can be chosen so that the matrix can be inverted. With the size and speed
of modern computers this is can even be accomplished with brute force
techniques. In many cases $A$ will be a multiple of a kernel which is the
result of a single observation.

Now to get a form of equation (12) suitable for computing, let
\begin{equation}
T=A^{-1/2}BC^{-1/2}=(H^i_q H_{ir})^{-1/2}H^j_r H_{j\alpha}(H^k_\alpha 
H_{k\beta})^{-1/2}.
\end{equation}
Then
\begin{equation}
X = (A-BC^{-1}B^T)^{-1}Y
  =A^{-T/2}(I-A^{-1/2}BC^{-1}B^TA^{-T/2})^{-1}A^{-1/2}Y
  =A^{-T/2}(I-TT^T)^{-1}A^{-1/2}Y.
\end{equation}
Like $B$, the size of $T$ is ${\cal P} \times \Gamma$, but it is sparse.

Finally, we use $(I-TT^T)^{-1}=\sum^\infty_{n=0} (TT^T)^n $ to get a form
that is tractable with a modest computer. Although formally the sum
must be carried to infinity the sum converges in tens to hundreds of
iterations for well chosen observations. Then,
\begin{equation}
X= QY=A^{-T/2} \left[ \sum^\infty_{n=0} (TT^T)^n \right] A^{-1/2} Y.
\end{equation}

The matrix, $TT^T$, is avoided by defining $Z_0=A^{-1/2}Y$, and iterating
\begin{equation}
Z_{n+1}=Z_0+T(T^TZ_n)
\end{equation}
until $Z$ is stable.
It is trivial then to get the solution $X=A^{-T/2}Z$. This is only the solution 
for the detector, but the solution for the sky is then straight forward.

\begin{deluxetable}{rl}
\tablewidth{0pt}
\tablecaption{Variable Definitions}
\tablehead{
\colhead{Variable} & 
\colhead{Definition}
}
\startdata
$P$ & number of Array pixels, e.g. $256\times256 = 65536$\\
${\cal P}$ & number of detector parameters, e.g. $2P = 131072$\\
$M$ & number of images in the data set, e.g. 100\\
$i,j,k$ & are indices to data $\in(1 \dots P\times M)$\\
$D^i$ & data\\ 
$\Delta^i$ & model error\\
$V^i$ & data variance (assumed to be diagonal)\\
$\Gamma$ & number of observed sky locations, e.g. 500000\\
$\alpha,\beta$ & indices to sky locations $ \in (1 \dots
 \Gamma ) ,\Gamma < P \times M)$\\
$S^\alpha$ & set of sky parameters\\
$p$ & index to pixels $ \in (1 \dots P)$\\
$G^p$ & set of gain parameters\\
$F^p$ & set of offset parameters\\
$q,r$ & indices to detector parameters $\in (1 \dots {\cal P})$\\
$X^q$ & set of  detector parameters $ (\delta F^p,\delta G^p)$\\
$\mu ,\nu,\omega$ & indices to all parameters $\in(1 \dots {\cal P}+\Gamma)$\\
\enddata
\end{deluxetable}

\section{Example}
Next we show how the algorithm is used in a practical program.
Some of the key details are given in the appendix, here we outline the steps
of the program and relate them to the previous derivation.

We adopt a simple model for the data, but more complex models
are as easily handled as long as they are relatively linear in the
range of interest, do not require large numbers of parameters to be determined,
and are not undetermined. A formal derivative must be calculated for each 
of the extra parameters and coded into the algorithm, while this may be messy 
and clutter the program, small numbers of parameters (e.g. temperature effects)
that affect the entire array make only small changes to the required time
or the final accuracy of the algorithm. If some part 
of the parameter space is undetermined the program may not converge.

Our example has a separate gain, $G$, and offset, $F$, for each detector
that modify the sky intensity, $S$, as it is detected. The model, $H^i$ for 
the data is given by
\begin{equation}
H^i(G^p,S^\alpha,F^p)= G^p S^\alpha + F^p.
\end{equation}
\begin{equation}
X=(\delta G^1\dots \delta G^P,\delta F^1\dots \delta F^P)
\end{equation}
The example is obviously nonlinear and we must be careful to chose an initial 
point close enough to the solution for the algorithm to converge to the 
solution. For a particular detector array one would use the algorithm
many times so one can use the last solution as the starting point and
either add more data to improve the solution or find a new solution with new
data. Either way, only once, do we need to start without a previous solution.
In that case we can let $G^p=1$, $F^p=0$, and $V^p=\sum_{i\in p} (D^i-F^p)^2/M$.
Then with the assumption that the uncertainties are a function of pixel
only, we have an estimate for $V^i$. We will return to this estimation in
section 6.

We assume a diagonal weight matrix $W_{ii}=(V^i)^{-1}$ to keep the
example simple. However, we emphasize that this is {\it not} required. The 
derivation is completely general and can accommodate a nondiagonal weight
matrix. Note that this assumption does not
mean that the data are uncorrelated. Indeed, the data are correlated as
some of the data are derived from the same pixel or are observations of the 
same part of the sky with different detectors. If there are other sources of 
correlation (such as detector temperature) they need to be explicitly included 
in the model. The assumption here is that the residual errors are uncorrelated.

The first step of the program is to calculate
\begin{equation}
\Delta_i=W_i(D^i-G^pS^\alpha-F^p).
\end{equation}

As there are two types of parameters we divide the matrix $A$ into its four
quadrants for discussion.
\begin{equation}
A=\left(
\begin{array}{ll}
 A_G & A_{GF} \\
 A^T_{GF} & A_F
\end{array}
\right).
\end{equation}
Each of the submatrices of $A$ is diagonal, including the part relating the 
gain and offset of each pixel. The whole matrix is treated as $P$ $2\times2$ 
matrices. There is not a unique $A^{-1/2}$, mathematically the choice is 
arbitrary, but the symmetric choice and the 
choice where the lower left are zero are easier to program. We have used both
and found the nonsymmetric version is less susceptible to numerical instability.

Although the size of $H^i_\mu$ is $PM\times({\cal P}+\Gamma)$ it can be treated as a set of
delta functions. With care in the processing, the parts of $H$ that are zero 
need never be accessed (appendix). There are 3$P\times M$ nonzero parts. 
That is each datum appears 3 times, once associated with $G$, $F$ and $S$.

The second step makes use of the following relations:
\begin{tabular}{l}
$\partial_{G^p}H^i = S^\alpha\delta_{pi}$ \\
$\partial_{F^p}H^i = \delta_{pi}$\\
$\partial_{S^\alpha}H^i = G^p \delta_{\alpha i}$\\
\end{tabular}
to construct

\begin{equation}
{\rm diag}~~A_G = \sum_{i\in p,i\in \alpha} S^\alpha W_i S^\alpha,
~~~~~{\rm diag}~~A_F =\sum_{i\in p} W_i
\end{equation}

\begin{equation}
{\rm diag}~~A_{FG} =\sum_{i\in p,i\in \alpha} S^\alpha W_i
\end{equation}
and
\begin{equation}
{\rm diag}~~C=\sum_{i\in \alpha,i\in p} G^p W_i G^p.
\end{equation}
$C$ is diagonal as well.

The matrix $B$ is divided into two parts similar to $A$:
\begin{equation}
B_G=\sum_{i \in p,i \in \alpha} S^\alpha W_i G^p,
~~~B_F=\sum_{i\in p,i\in\alpha} W_i G^p.
\end{equation}

Finally $Y$ has two parts
\begin{equation}
Y_G= \sum_{i\in \alpha,i\in p} S^\alpha \Delta_i
-B_GC^{-1}\sum_{i\in p,i\in \alpha} G^p\Delta_i,
~~~Y_F=\sum_{i\in p} \Delta_i-B_FC^{-1}\sum_{i\in p,i\in \alpha} G^p\Delta_i.
\end{equation}

Note that $B$ is $2P\times\Gamma$ but it is sparse. We then calculate 
$T=A^{-1/2}BC^{-1/2}$. With the elements of equation (16), the program iterates 
equation (17) until $Z$ is stable. Then the solution 
$X=A^{-T/2}Z$. This is only the solution 
for the detector, but the solution for the sky is then:
\begin{equation}
S^\alpha=\sum_{i\in\alpha, i\in p} [(D^i-F^p)G^pW_i]
 /\sum_{i\in\alpha, i\in p}(G^p)^2W_i.
\end{equation}

This then is a form which can be handled by a modest computer. The vectors
$X$ and $Y$ are each only $2P=131072$ long. The matrix $A$ is stored as
three $P$ long diagonal parts of its submatrices. The matrix $T$ is nominally
large ($2P\times \Gamma$) but is sparse and has at most $2P\times M$ nonzero
components.

At this point there are two obvious singularities.
These correspond to the uniform change in the sky brightness and a cancelling 
change in the offset, and to a multiplication of the sky by an arbitrary 
amount and a cancelling effect in the gain term. These two singularities 
point out what we already know; in order to get an {\it absolute} calibration 
we need an {\it absolute} standard. There are several ways to deal with this 
issue: 1) An absolute calibration could be done in the laboratory. 2) Certain 
places on the sky could be determined in some other way and used to impose a 
condition that would break the singularity.
3) A map could be produced with an arbitrary gain and offset.

The three methods are not mutually exclusive. A map with arbitrary gain and 
offset can be produced which is subsequently calibrated by laboratory 
measurements or sky measurements or a combination of sky and laboratory
measurements. The absolute calibration can be included in the fit or applied 
later. We choose to apply it separately as this maintains the uniformity of the
algorithm whether viewing a calibration object or not.

Without treating the singularity, the sum in equation (16) does not converge.
If there are no dark frames to determine the offset, after each iteration we 
impose the condition that $\sum_p \sqrt{\sum_{i\in p}W_i}~\delta F^p =0$.
The weight applied to the $\delta F^p$ is only for computational convenience
(it is the form of $F^p$ in $Z$). The key is that the net offset is not allowed 
to change.
If dark frames are present we can use them to determine the offset and do
not impose this condition. Similarly, a weighted mean gain is held fixed, 
$\sum_p \sqrt{\sum_{i\in p, i\in\alpha}S^\alpha W_i S^\alpha}~\delta G^p=0$.

This completes the solution for the detector and the sky. The calculation
of a single uncertainty vector is completely analogous. However the full
covariance matrix $\Psi$ is $\sim500000\times500000$. This matrix is symmetric
but it is not sparse. In fact it is likely that all of the
elements are nonzero. The $2.5\times10^{11}$ components of $\Psi$ are awkward 
to carry around but they contain all of the information about the
correlations imposed by the calibration process. It can be stored more
compactly by keeping $T$, and noting that
\begin{equation}
\Psi= C^{-1/2}(I-T^TT)^{-1}C^{-1/2}
\end{equation}
since $T$ is sparse and $C^{-1/2}$ is diagonal.

Now we return to the issue of variance (weight) estimation. Without a
model for the noise we have a hopeless task. However with a simple model
we can estimate the variance. An unbiased, but poor,
estimation only increases the noise (and the estimate of the uncertainty).

In the model program we assume three sources of error: 1)Poisson statistics, 
2)A pixel dependent readout noise, 3)A cosmic ray induced error. The Poisson
noise is easily calculated if the approximate gain of the system is known. 
The readout noise is best
estimated by using the RMS of all of the data from that pixel (except the 
cosmic ray contaminated data). Cosmic rays are identified by seeking
large discrepancies. These should not be used in either the sky or variance
estimation. Obviously as data are collected a more detailed model
can be developed.

After a solution is found, the model program recalculates $\Delta^i$. 
Data with errors greater than $2.5~\sigma$
are assumed to be cosmic particle hits or other glitches. These data
are marked and not used in the next iteration. The remaining $\Delta^i$s
are squared and summed to estimate the noise. The model program noise is 
treated separately for each pixel. If hundreds or thousands of pictures
are available this process could potentially identify subtle problems with
particular pixels. If fewer data are available a smooth
approximation over entire detector array is more appropriate.

\section{Practical Matters}

The algorithm described in the preceding sections can produce mosaics 
of large regions provided that at least some parts of the region
(preferably all parts) contain repeated (dithered) observations. 
The algorithm can be applied to a data set containing spatially 
separate regions. There is no constraint on the size or geometry of the
region(s) in the data set. It is only required that the detector gain and
offset and the sky intensity ($G^p$, $F^p$, and $S^{\alpha}$) are constant
for the entire data set. These restrictions can be relaxed by explicitly
parameterizing known or suspected variations.

If dark frames are available, they are added to the data set as if
they are observations of a region of sky that is separate from the 
rest of the data and that has an intensity $S^{\alpha} \equiv 0$.
The addition of dark frames to the data set allows the algorithm 
to determine the offset components.

The algorithm can be implemented in a general manner, such that the detector 
dimensions and number of frames processed are adjustable. A general code can
be applied to different data sets from different instruments if a new
``front end'' is written for each type of data to ingest the data and
provide the necessary initial estimates and control parameters.

The selection of the weights ($W_{ii}$) to use in the algorithm can be 
important. Poor weighting of the data may cause spurious features to
propagate through the solutions for $G^p$, $F^p$, and $S^{\alpha}$.
Cosmic ray hits on the detector also cause spurious features
in the results, if not properly handled. Data affected by cosmic rays
can be given very low weights or flagged. It is best if the effects of
cosmic rays are removed from the data before processing, though this
is not always possible. The algorithm can recognize 
cosmic rays as outliers provided that they are not so numerous that they
severly bias the results.

In most cases, the algorithm will be used iteratively for 
2 - 5 cycles. Subsequent iterations use the previously derived gain and 
offset values as inputs, and make use of successively improved weights 
and exclusions of cosmic rays as well. 

An IDL implementation of this algorithm requires free memory $\sim$15 times 
larger than the size of the data set to be processed. For a data set of
27 256$\times$256 images the algorithm takes $\sim$450 seconds of CPU time on
a 300 MHz Pentium II machine running Red Hat Linux 5.2 and IDL 5.0.3.
About 270 seconds of that CPU time is spent in the calculation of the 
summation of equation (16), using the iterative step of equation (17) for 100 
terms. The key data arangements of the program are discussed in the appendix.
The time for the procedure is linear with the number of input data 
elements as long as more iterations are not needed. The number of iterations
required is strongly related to the connection map which is determined
by the dither pattern of the input data.

Solving only for detector gains in cases where the detector offsets are
negligible is a minor simplification of the algorithm and is a more
robust procedure. Figure 1 illustrates the results of using this procedure to 
solve for only the detector gains and sky intensities. The data used is 
from Wide-Field Infrared Explorer (WIRE) simulations. The model for the sky
includes point sources, cirrus and a zodiacal background. The model for the 
$128\times128$ detector array included gain variations, bad pixels, and 
cosmic ray hits. The data
set consists of 10 dithered images, one of which is shown in the upper left 
of Figure 1. The detector gain variations dominate the qualitative appearance
of the data. The derived gain compares favorably with the true gain, with the
exception of $\sim0.2\%$ of the pixels with remaining artifacts from bad 
detectors and cosmic
rays. There is a small ($\sim1.0025$) scale factor between the derived and 
true gains, which reflects the lack of absolute calibration in the 
procedure. The derived sky is a good representation of the real sky, with the 
additional noise component indicated by the second term of equation (11). 

Figure 2 illustrates the application of the algorithm to real data, namely
the HST NICMOS observations of the Hubble Deep Field - South. The raw data used
here were 59 good 1152 and 1472 s integrations. The worst effects of cosmic 
rays were eliminated by calculating linear fits to the multiaccum readouts 
from each pixel. Fits with poor correlation coefficients were refit using
a combination of linear and step functions. Additional pre-processing involved
subtracting the median value of each quadrant of each frame from that frame
quadrant. This helped compensate for a variable ``pedestal'' effect which is not 
modelled by our current algorithm. (The bottom 16 rows of each frame were
ignored in the processing to avoid vignetting effects.) The initial gain map 
was assumed to be flat and unity. The initial offset map was assumed to be flat 
and zero. A dark file from the NICMOS reference files was used for a simulated 
dark frame that was processed simultaneously with the sky data.
The derived sky after 2 iterations of the algorithm and truncation of the series
expansion after 100 terms, is not as clean as the publicly released 
processed data. Spurious large scale structure is present at low levels.
A faint stripe along the detector columns is visible through the brightest
star in the field. The gain and offset maps are similar to calibration flat 
and dark reference files. In our derived gain and offset maps there are 
residual defects in pixels where the bright sources in the map were observed. 
The gain and offset maps also contain visible quadrant errors and vertical
bars from ``shading" because of instrumental effects that are not adequately
described by the simple method used here.
Clearly there is room for improvement, but the algorithm 
worked well. The process allowed the simultaneous determination of sky and
detector parameters using only sky measurements and dark frames.
By inspecting the residuals there are indications that the offsets are
not constant from observation to observation. This suggests an improved 
model for the data could be constructed by parameterizing and fitting
these offsets.

In the case of IRAC, such an algorithm is essential. With it, we can continuously
derive detector parameters from the normal observations and improve the 
model of the detectors as well as the model of the sky. Just such a procedure
was used on the FIRAS data to improve the sensitivity by a factor of
$\sim100$ over the initial publication.

\section{Uncertainties and Correlations}

The algorithm produces a formal estimate of the uncertainties, $\Psi$, based on the 
derivation and the estimated uncertainties of the input data, $V$. The resulting
uncertainties are only as good as the uncertainty estimates of the original
data. Those uncertainties, $V$, are checked against the actual deviations from 
the model to either give an improved estimate of the input data uncertainty or 
an indication of short comings of the model.

Identifying the weight matrix (or metric) as the inverse of the covariance
matrix, only defers the question to how to determine the covariance matrix.
There are two sorts of ways to attack this problem. The theoretical approach
uses {\it a priori} knowledge about the system to estimate what the noise should be.
This includes such things as the Poisson arrival of photons, the Johnson
noise of the resistors and other known sources of noise. The empirical
approach uses the residuals in the data itself to make an estimate of the 
noise. Each approach has its strengths and weaknesses. The theoretical 
approach often underestimates the noise because there are unmodeled noise sources
present. The empirical approach often overestimates the noise, as
it treats parts of the signal that are not properly modeled as noise. If
both approaches lead to the same estimate one has reasonable assurance that
the model and estimate are correct. If the approaches differ significantly
there are either noise sources that are included in the estimate or signal 
that is not included in the model. In this example, we 
assume that the noise variance $V$ is known.

The calibration process introduces correlations into the resulting map,
$C^{-1}B^TQBC^{-1}$. The correlations for a single detector are easily 
generated by using a unit vector in place of the $Y$ in equation (12) and 
carrying out the 
calculations as with the data. The process is slightly shorter than for the 
data (checking for convergence is omitted). Obviously this could be repeated for each 
of the detectors and then equation (11) could be used to generate the full 
covariance matrix.

There are two problems with this approach. First, the time required is 
proportional to the number of detectors (65536 for IRAC or NICMOS data). 
Second, the space for the final result is $\Gamma^2$, which is $\sim10^{11}$,
for even the modest WIRE example shown here. Storing and using such a large
data set is problematical.

Fortunately the correlations for different detectors are nearly identical
(see figure 3). This should not be a surprise since the detectors are locked
into their relative positions and all move together in each dither move.
Bad detectors in the array, cosmic rays, and rotations will obviously
break this symmetry but except for the rotations the effects are minor
and rather localized. So the correlations can be calculated for typical
detectors and the results can be used for the entire data set.

\section{Summary}

As demonstrated with the simulated WIRE data the program can calculate the 
gains and offsets to the theoretical limit on the accuracy if it is given
a good model of the data. As shown with the NICMOS data the program works
reasonably well on real data as well even with he normal complexities of
real errors and uncertainties. The uncertainties are calculated, and the 
correlations can be calculated with minor changes to the program. These allow
the user to interpret the result without ad hoc assumptions or guesses about
how the errors are related. The speed of the program allows modest data
sets to be processed in a few minutes and with the availability of machines
with large memories will allow the large data sets of the future to be processed
in reasonable times.

\section{Acknowledgements}
We thank D. Shupe and the WIRE team for supplying simulated WIRE observations
for testing the algorithm.

\section{References}
\frenchspacing
\setlength{\leftmargini}{0cm}
\begin{verse}

\bibitem[arendt 00]{dither}{Arendt, R.~G., Fixsen, D.~J. \& Moseley, S.~H. 
2000 ApJ submitted}

\bibitem[Fazio 98]{irac}{Fazio, G.~G. \etal 1998, Proc. SPIE, 3354, 1024}

\bibitem[Fixsen 94]{Calib}{Fixsen, D.~J. \etal 1994 ApJ 420,427}

\bibitem[Fruc]{}{Fruchter, A.S. \& Hook R. N. 1998, 
preprint (astro-ph/9808087)}

\bibitem[Joyce]{}{Joyce, R. R., 1992 ASP Conference Series, Vol. 23, Astronomical CCD
Observing and Reduction Techniques, ed. S. B. Howell (San Fransisco: ASP)}

\bibitem[NumR]{Num}{Press, W. H., Teukolsky, S. A., Vetterling, W. T., 
\& Flannery, B. P. 1992, Numerical Recipes in C: The Art of Scientific 
Computing, Second Edition, Cambridge: Cambridge University Press}

\end{verse}
\nonfrenchspacing

\appendix{}
\section{Code Considerations}

This appendix points out several details of implementing this
least squares calibration procedure in a computer code.
The first detail is that a sparse matrix storage and multiplication
system must be applied.
As presented here, the solution (eq. 16) requires construction
of the matrix $T$ which has dimensions $\Gamma\times{\cal P}$.
For a $256\times 256$ detector array, $T$ contains at least $256^4 =
17\times 10^9$ elements, making it difficult to store in memory
However most of the elements of $T$ are 0.0, because a single detector,
$p$, observes no more than $M$ of the $\Gamma$ sky pixels.
Following the example presented in \S4, we note that $T$ contains
the same non-zero elements as $B$. Furthermore, each datum
leads to one element in $B_G$ and one element in $B_F$ (eq. [25]).
Thus $B$ or $T$ can be stored in an array corresponding to the $P\times M$
data, and replicated for each of the detector parameters (gain, offset, etc.)
to be determined. The position within the array indicates the
$p \in [1,{\cal P}]$ index of the element, while the $\alpha \in [1,\Gamma]$
index is stored in a separately constructed array. In this way, the
storage requirements are reduced by a factor of $\sim({\cal P}/P)M/\Gamma$
which is generally very large as $M\ll\Gamma$ for most datasets.

The second detail is to note that equation (17) is can be implemented as
a pair of matrix $\times$ vector multiplications: $T^TZ_n$, followed by
$T(T^TZ_n)$. This pair of multiplications is much faster and requires
negligible storage compared to calculating the matrix multiplication
$TT^T$ first, and then $(TT^T)Z_n$. The $TT^T$ matrix is not nearly
as compact as the $T$ matrix. Furthermore, with the appropriate juggling of
indices both matrix $\times$ vector multiplications are performed using
the stored format of $T$ without explicitly calculating the transpose of $T$.
An example of this is found in Press et al. (Chapter 2, 1992).

\clearpage
\figcaption[wire]{The top left image shows one of ten frames of simulated WIRE data.
The top center image shows the detector gains derived from the data, while the 
top right image shows the actual gains used to generate the simulated data.
The lower left graph shows the histogram of the differences between the derived 
and actual gains. The bottom middle and right images show the derived sky 
intensities and the true sky used to generate the simulated data.}

\figcaption[nicmos]{The raw data is one of the NICMOS multiaccum frames after 
fitting linear fits to the readouts from each pixel and removing the worst
of the cosmic rays. The other pairs of derived and reference images are 
each shown on equivalent scales. The derived gain and offset maps only cover
the upper $256\times240$ detectors in the $256\times256$ array.}

\figcaption[cor]{The panels show six columns of the $2P\times 2P$ matrix 
$A^{T/2}QA^{1/2}$ for a $256\times 256$ detector array and an idealized data 
set collected using a dither pattern consisting of 36 pointings evenly spaced 
along the sides of a Reuleaux triangle. The columns are reformated into 
$256\times (256*2)$ arrays. From left to right and top to bottom the columns 
correspond to those containing the correlations for $G^p$ $(p = [128,128],
[16,128],[16,16])$ and $F^p$ $(p = [128,128],[16,128],[16,16])$. Correlations 
against $G^p$ and $F^p$ map into the bottom and top half, respectively, of 
each panel. Black indicates strong positive correlations. Displayed ranges for 
$G^pG^p$, $F^pF^p$, and $G^pF^p\ =\ F^pG^p$ correlations are 
[$1.5\ 10^{-3}$, $1.55\ 10^{-3}$], [$1.2\ 
10^{-3}$, $2.0\ 10^{-3}$], and [$-8\ 10^{-5}$, $8\ 10^{-5}$] respectively.}

\end{document}